\documentstyle[12pt]{article}

\newcommand{\del}{\partial}
\newcommand{\df}{\Delta f}
\newcommand{\dfij}{\Delta f_{ij}}
\newcommand{\hamlij}{2 E |\phi(r)| \dfij}

\newcommand{\hamll}{E |\phi(r)| \df}
\newcommand{\ket}{\rangle}
\newcommand{\thg}{\theta_{G}}
\newcommand{\temu}{{\theta}_{12}}
\newcommand{\tetau}{\theta_{13}}

\newcommand{\ra}{\rightarrow}
\input epsf
\begin{document}
\begin{titlepage}
\begin{flushright}
  WATPHYS TH-95/01\\
  hep-ph/yymmddd\\
\end{flushright}
\vspace{0.25in}
\begin{center}
{\large\bf Adiabatic Analysis of Gravitationally-Induced Three-Flavor
Neutrino Oscillations}\\
\vspace{10pt}
J.\ R.\ Mureika \\
{\it Department of Physics} \\
{\it University of Waterloo} \\
{\it Waterloo, Ontario N2L 3G1  Canada} \\
{[email: newt@avatar.uwaterloo.ca]} \\
\vspace{13pt}
R.\ B.\ Mann\footnote{on leave from Physics. Dept.,
 University of Waterloo, Ontario, Canada}\\
{\it Department of Applied Mathematics and Theoretical Physics}\\
{\it Silver Street}\\
{\it Cambridge University}\\
{\it Cambridge U.K.  CB2 9EW}\\
{[email: rbm20@damtp.cam.ac.uk]}\\
\end{center}
\vspace{15 pt}
\begin{center}
{\bf Abstract}
\end{center}
Some implications of
the proposal that flavor nondiagonal
couplings of neutrinos to gravity might resolve the solar neutrino
problem are considered in the context of three neutrino flavours.
Using an adiabatic approximation iso-SNU curves are calculated for the
neutrinos which most
likely contribute to the observed solar neutrino deficiency.  We show
that results obtained with two-flavor models can be recovered, and
discuss the effects of the addition of a third flavor.  Some
results are obtained for this case, and these values are compared with
recent experimental data.
\end{titlepage}
\setcounter{page}{1}

\section{Introduction}

The Solar Neutrino Problem (SNP) is a mystery which mars our current
understanding of stellar astrophysics.  There are now four experiments
\cite{clx,galx,k2,sage} employing different detection techniques
that consistently find
a deficit in the observed $\nu_e$ flux relative to that predicted by
various solar models (see {\it e.g.} those listed in \cite{bah2}).
The origin of this discrepancy is unknown, but various possible solutions
based on particle physics techniques have been proposed.

	One possible class of solutions to this dilemma invokes
oscillations between the various flavors or generations of
neutrinos.  Perhaps the most elegant solution of this type proposed
to date is the Mikheyev-Smirnov-Wolfenstein (MSW)
Mechanism \cite{msw1,wolf}; here, the Hamiltonian becomes off-diagonal under
a change of eigenbasis for the neutrinos, resulting in flavor-oscillations,
much akin to those found in the quark sector.  However, the MSW effect
requires the neutrinos to possess a mass eigenbasis, and hence implies
massive neutrinos.

	In 1988, Gasperini \cite{gasp1} conjectured that a similar
oscillation phenomenon could take place, while still ensuring massless
neutrinos. In this mechanism each neutrino flavor is assumed to couple
differently to gravity,  permitting flavor oscillations at the expense
of violating the equivalence principle. This mechanism has been
investigated in a number of papers (see {\it e.g.} \cite{mal1,halp1,pant1}),
and has recently been referred to as the VEP mechanism \cite{bah1}.
The results of these studies indicate that, although strongly constrained by
existing data, the VEP mechanism is at present a phenomenologically
viable means of generating neutrino oscillations.

	All investigations of the VEP mechanism to date have restricted
consideration to 2-flavor oscillations. However, at our current level of
understanding, we know there to be at least three generations of neutrinos.
While two-flavor oscillation models are elegant and mathematically simple,
it remains a mystery whether or not the $\tau$-neutrino ($\nu_{\tau}$)
plays a crucial role in determining the observed $\nu$ counting rates
and fluxes.  Several studies  \cite{bald,barg,mal2,msw2,zag} have
shown that,
for certain cases, a three-flavor analysis can and does contribute
interesting physics to the SNP.

	In this paper we consider an adiabatic extension of the VEP
mechanism to the realistic three-flavor case.  The width of the parameter
space region found in \cite{bah1} is reproduced for a two-flavor
limit, and the size of the region is found to shift and/or widen for certain
values of the $\nu_e \ra \nu_{\tau}$ mixing angle.  Spectra for
the solar ${^8}B$ neutrinos are produced for differing values of the
mixing angles (for fixed violation parameters $\df_{21}$ and $\df_{31}$),
both for a fixed $^8B$ neutrino event rate, as well as for a fixed event
rate as would be detected by the $^{37}Cl$ experiment ({\it i.e.} combined
counting rate for ${^8}B + {^7}Be$ neutrinos).  These are
compared with the predicted flux in the two-flavor limit.

\section{Review of Three-Flavor MSW Neutrino Oscillations}

Before proceeding to the gravitational case, we review the MSW mechanism
for three flavors. Consider the neutrino eigenbases
\begin{equation}
|\nu_W \ket = \left( {\begin{array}{c} \nu_e \\ \nu_{\mu} \\ \nu_{\tau}
\end{array}} \right) ~,~|\nu_W \ket = R_3 \, |\nu_M \ket ~,
\end{equation}
where $|\nu_W \ket$ is the weak eigenbasis and $|\nu_M \ket$ the mass
eigenbasis. The matrix $R_3$ is the $3 \times 3$ leptonic analogue of the
CKM mixing matrix, which we parametrize as

\begin{equation}
R_3 = e^{i \psi \lambda_7} e^{-i \phi \lambda_5} e^{i \omega \lambda_2}~,
\label{ckm}
\end{equation}
and is proportional the the identity matrix in the case of massless neutrinos.
Here $\psi$, $\phi$, $\omega$ are three mixing angles, and $\lambda_k$ the
$SU(3)$ generators.  For simplicity we have taken $R_3$ to be real, which
implies that there are no CP violations in this sector ({\it i.e.} the
CP-violating phase $\delta=0$).  Including matter effects, and (without
loss of generality) assuming  $m_3 > m_2 > m_1$, the Hamiltonian in the
mass eigenbasis becomes
\begin{equation}
\begin{array}{ll}
H' = \nonumber \\
\frac{1}{2} \left( \begin{array}{ccc} m_1^2+m_2^2+A\,c_{\phi}^2-
\Delta_N & 0 & A\, s_{2 \omega} \sqrt{(1- \beta)/2} \\ 0 & m_1^2+m_2^2+
A\,c_{\phi}^2+
\Delta_N & A\, s_{2 \omega} \sqrt{(1+\beta)/2} \\ A\, s_{2 \omega} \sqrt{(1-
\beta)/2} & A\, s_{2 \omega} \sqrt{(1+\beta)/2} & 2 m_3^2 + 2 A\, s_{\phi}^2
\end{array} \right)
\end{array}
\label{mix}~.
\end{equation}
where  $c_{\phi} = \cos \phi$, {\it etc}. We follow the definitions of
\cite{bald} with
\begin{eqnarray}
\Delta_N & = & [(A\, c_{\phi}^2 - \Delta_{\mu}\, c_{2 \omega})^2
+\Delta_{\mu}^2
s\,_{2 \omega}^2]^{1/2} \\
\beta & = & \frac{1}{\Delta_N} (A\, c_{\phi}^2 - \Delta_{\mu}\, c_{2 \omega})^2
\end{eqnarray}~
where  $\Delta_{\mu} \equiv m_2^2 - m_1^2$, and $A=A(r)$ is the matter
interaction term,
\begin{equation}
A(r) = 2 \sqrt{2} \: G_F N_e(r) E~.
\end{equation}
which describes the interaction of $\nu_e$-neutrinos with solar matter.

	From this point on, we make the following substitutions for the
mixing angles of  (\ref{ckm}): $\omega = \temu$, $\phi = \tetau$, $\psi =
\theta_{23}$. Assuming that the the state transitions are adiabatic,
we can approximate the $\nu_e$ survival probability
$P_{\nu_e \ra \nu_e}(r,E)$ to be
\begin{eqnarray}
P_{\nu_e \ra \nu_e}(r,E) & = & (1-P_{LZ2}) \, \Theta(A-\Delta_{21})
+ P_{LZ2}(1-P_{LZ3}) \, \Theta(\Delta_{21}-A) \nonumber \\
			 &   & \mbox{} \times
\Theta(A- \Delta_{31})+P_{LZ1} P_{LZ3} \, \Theta( \Delta_{31} - A)~ .
\label{probs}
\end{eqnarray}
Here $\Theta(x)$ is a step function
\footnote{The step function is a good approximation for the sines and
cosines of the matter-enhanced mixing angles, in the case of small
vacuum angles \cite{bald}.  The results obtained using this form of
the survival probability are in excellent agreement with the full
form of the (adiabatic) probability.}, with
\begin{equation}
\Theta(x) = \left\{ \begin{array}{ll} 1 & \mbox{$\forall x<0$} \\
				    0 & \mbox{otherwise} \end{array}
\right. \, ~,
\end{equation}
and $P_{LZi}$ is the Laudau-Zeener ``jump" probability for the
$\nu_e \ra \nu_i$ transition,
\begin{equation}
P_{LZi} = \frac{e^{-\beta_i} - e^{-\alpha_i}}{1 - e^{-\alpha_i}} ~ ,
\end{equation}
with
\begin{equation}
\alpha_i = 2 \pi \kappa_{i} \left( \frac{\cos 2 \theta_{1i}}{\sin^2 2
\theta_{1i}}
\right) ~,~ \beta_i = \frac{\pi}{2} \kappa_{i} (1- \tan^2 \theta_{1i}) ~.
\end{equation}
For $\nu_{e} \ra \nu_{\mu}$ transitions, $i=2$, and $i=3$ for $\nu_e \ra
\nu_{\tau}$.

\section{Gravitationally-Induced Neutrino Oscillations}

We now modify the equations of the previous section to model neutrino
oscillations induced by the non-zero difference $\dfij$ between the
gravitational
coupling of the different flavors.  Consider first the 2-flavor scenario.
Let $|\nu_W \ket$, $|\nu_G \ket$ be the respective weak and gravitational
eigenbases
\begin{equation}
|\nu_W \ket = \left(\begin{array}{c} \nu_e \\ \nu_{\mu} \end{array} \right) ~,~
|\nu_G \ket = \left(\begin{array}{c} \nu_{1G} \\ \nu_{2G} \end{array} \right) ,
\end{equation}
where the two are related by an $SO(2)$ rotation $R \equiv R(\thg)$
(the ``mixing" matrix),
\begin{equation}
|\nu_W \ket = R \, |\nu_G \ket ~ .
\end{equation}

The gravitational eigenstates obey a variation of the Dirac Equation
\begin{equation}
[i({e_a}^{\mu} \gamma^a \del_{\mu}) - \zeta(\phi)] |\nu_G \ket = 0 ~,
\label{dirac}
\end{equation}
where $\zeta(\phi)$ is some function of the gravitational potential, which
arises due to the non-zero violation of the equivalence principle.
For a spherically symmetric metric (choosing angular coordinates $\theta =
\phi = 0$ as the trajectory of the neutrino), the vierbeins
$e^a_{\mu}$ in (\ref{dirac}) reduce the evolution equation to \cite{min}
\begin{equation}
i \frac{d}{dr} |\nu_G \ket = H \, |\nu_G \ket ~ .
\end{equation}
where
\begin{equation}
H_i = -2 |\phi(r)| E_i (1+f_i) ~  .
\label{hamil}
\end{equation}
are the diagonal components of the Hamiltonian in the gravitational
eigenbasis \cite{pant1}.  Here, $|f_i| \ll 1$ is a (flavor-dependent)
parameter which determines the magnitude of the violation of the Einstein
Equivalence Principle (EEP).  Note that for $\nu_e$ (i.e. first
generation matter), $f=0$, which differs from the notation of
\cite{bah1} (in which $f=1$ represents first-generation neutrinos,
which is equivalent to saying that the term $(1+f) = 1$ in
(\ref{hamil})).  Under the change of basis $|\nu_G \ket \ra
|\nu_W \ket$, the Hamiltonian becomes off-diagonal,
\begin{equation}
i \frac{d}{dr} |\nu_W \ket = R^{-1} H R \, |\nu_W \ket ~ ,
\end{equation}
and with the inclusion of weak interactions in the $\nu_e$ sector,
we obtain the effective Hamiltonian
\begin{eqnarray}
H^{\prime} & = & R^{-1} H R \: + \: W_e \\
  & = & \quad \left( \begin{array}{cc}
{\frac{\sqrt{2}}{2}\: G_F N_e(r)-\hamll \cos{2 \thg}} &
\hamll \sin{2 \thg} \\
\hamll \sin{2 \thg} & 0 \\ \end{array} \right)\nonumber
\end{eqnarray}

These matter-enhanced oscillations are resonant
for complete flavor conversion, which implies the energy condition
\begin{equation}
E = \frac{\sqrt{2}\: G_F N_e(r)}{2 |\phi(r)| \df \cos {2 \thg}}
\label{energy}  ~,
\end{equation}
where $G_F$ is the Fermi Coupling Constant, $N_e(r)$ the radial
electron number-density, and $\phi(r)$ the dimensionless (solar)
gravitational potential.  Also, $\df \equiv f_2-f_1$.
In addition, we require the {\it adiabaticity condition}
\begin{equation}
\kappa =\frac{\sqrt{2}\: G_F\, (N_e)_{res}\: {\tan}^2{2 \thg}}{{\left|
\frac{1}{N_e}
\frac{dN_e}{dr} - \frac{1}{\phi} \frac{d \phi}{dr} \right|}_{res}} \gg 1 ~ .
\label{adia}
\end{equation}

Iso-SNU curves ({\it i.e.} curves of constant neutrino-counting rates) for
2-flavor gravitationally-induced neutrino oscillations
have been carried out in \cite{mal1}, \cite{pant1}, and later repeated
in \cite{bah1}. Two distinct regions of overlap in parameter space were
discovered for the given experimental data \cite{bah1}, namely a highly
non-adiabatic region (small mixing region), and an adiabatic
``large mixing" region.  However the authors in \cite{mal1} have
argued that these
regions are too small, statistically speaking, to be kept as viable
solutions to the SNP for equivalence principle violating neutrino flavors.

\section{Extension of VEP Mechanism to Three Flavors}

Turning now to the three flavor case, we substitute
\begin{equation}
\frac{1}{2E} \left( \begin{array}{ccc} m_1^2 & 0 & 0 \\
0 & m_2^2 & 0 \\ 0 & 0 & m_3^2 \end{array} \right) ~\ra ~
2 E |\phi(r)| \left( \begin{array}{ccc} f_1 & 0 & 0 \\ 0
& f_2 & 0 \\ 0 & 0 & f_3 \end{array} \right)~,
\end{equation}
which amounts to replacing the mass-squared difference in the
probabilities (\ref{probs}) with an EEP violation term,
\begin{equation}
\frac{\Delta m^2_{ij}}{2E} ~ \ra ~ \hamlij ~,
\end{equation}
where $i,j \in [1,3]$, $i \neq j$.  For the $\nu_e \ra \nu_e$
survival probability, just the $i=2,3$, $j=1$ terms come into play,
and we need only concern ourselves with the mixing angles
$\temu, \tetau$.

	Of course the entire parameter space $(\df_{21},\df_{31},
\temu,\tetau)$ is not fair game for this analysis.  In particular,
we must limit the space to only the areas involving resonances,
and which are allowed by adiabatic processes.  The physical
boundaries on the $\df$ parameter are easily calculable, and have in
the two-flavour case
been found to be \cite{bah1} $\df \in (10^{-18}, 10^{-12})$.
The same range holds for the parameters $\df_{ij}$.
This implies that a $\nu_e$
with $|E \df| \geq 10^{-12}$ MeV will not undergo resonance, and will
emerge from the sun unscathed, retaining its identity as a $\nu_e$.
The other $\df$ boundary is similar, but for the lower resonance.

	This determination of resonant behavior provides us with valuable
information.  There are now three distinct regions of parameter space which
are of physical interest:

\begin{itemize}
\begin{enumerate}

\item \underline{$\df_{31} > 10^{-12}, \df_{21} \in (10^{-18},10^{-12})$:}
\begin{itemize}
\item {Here, we should expect large dependence on $\nu_{\mu} \ra \nu_e$
transitions ({\it i.e.} $\df_{21}, \temu$), as the $\df_{31}$ parameter
puts all transitions to $\nu_{\tau}$ above the upper resonance.
In the limit $\tetau \ra 0$, we should expect to recover the two-flavor
EEP violation model.}
\end{itemize}
\item \underline{$\df_{31} \in ((10^{-18},10^{-12}), \df_{21} < 10^{-18}$}:
\begin{itemize}
\item {This range represents the opposite of the previous one, where the
$\nu_e \ra \nu_{\mu}$ transitions are not resonant.  Hence, we expect
to see a two-flavor limit with strong dependence on the $\nu_e
\ra \nu_{\tau}$ transition parameters ($\tetau, \df_{31}$).}
\end{itemize}
\item \underline{$10^{-18} < \df_{21} < \df_{31} < 10^{-12}$}:
\begin{itemize}
\item {The most interesting behavior is expected to occur here,where
both parameters are sandwiched into the allowed $\df$ parameter range.
Similar to the mass MSW mechanism \cite{zag}, we should expect to
see dependence on all four parameters in question.}
\end{itemize}
\end{enumerate}
\end{itemize}

The following section summarizes our analysis of the physically
relevant cases $\#1$ and $\#3$.

\section{Results}


	Using the neutrino spectra and solar mass data from \cite{bah2},
as well as the ${^{37}} \! Cl$ and ${^{71}} \! Ga$ cross section data from
\cite{bah3}, we have computed iso-SNU rates for case $\# 1$ as listed above.
The counting rate $C_{x}$ for a given detector material $x$ with energy
threshold $E_{min}$ and absorption cross-section $\sigma_x(E)$ is
\begin{equation}
C_x = \int_{0}^{R_{\odot}} \! dr \: r^2 \xi(r) \int_{E_{min}}^{E_{max}}
\! dE \: \phi(E) \sigma_x(E) P_{\nu_e \ra \nu_e}(r,E)~,
\label{count}
\end{equation}
for neutrinos with energy spectrum $\phi(E)$ and maximum energy $E_{max}$.
The function $\xi(r)$ represents the fraction of neutrinos of a given
type produced at radius $r$.  The rates obtained by numerical intergration
of (\ref{count})
are essentially commensurate with those obtained in ref.\cite{mal1}
and others ({\it e.g.} see \cite{bah1}) for the "large mixing region"
(to within a few percent).  The neutrinos which were assumed to contribute
the largest counting rates for each type of detector include, in order
of abundance:

\begin{itemize}
\item $^{37}\!Cl ~\ra~ {^8}\! B,\ {^7}\! Be$
\item $^{71}\!Ga ~\ra~ pp,\ {^7}\! Be,\ {^8}\! B$ ~.
\end{itemize}

Note a slight discrepancy in the upper limit of the region as
given in \cite{bah1}.
It is believed that any discrepancies could well
be the result of numerical errors during calculation, or from minute variations
in data for the specific neutrinos.  We show in Table 1 the ($3 \sigma$)
adiabatic edges of the parameter space in question, according to the rates
as given in \cite{bah1}:

\begin{eqnarray}
C_{Home.} & = & 2.55 \pm 0.25 ~ SNU \nonumber \\
C_{GALLEX} & = & 79 \pm 11.7 ~ SNU \nonumber \\
C_{SAGE} & = & 73 \pm 19.3  ~ SNU \nonumber ~.
\end{eqnarray}

The edges of the parameter space region for the Homestake ${^{37}}\!Cl$
detector
are roughly equal to the bounds in question, while the
$3 \sigma$ range for the ${^{71}}\!Ga$ experiments is much larger.  Hence,
the overlap range is determined predominantly by the ${^{37}}\!Cl$ counting
rates,
which have been calculated to be $\sin^2 2 \temu \in (0.60,\ 0.89)$.  This
is found to be in close agreement to the range width obtained by \cite{bah1}
of $\sin^2 2 \temu \in (0.60,\ 0.90)$.

	As previously mentioned, the most interesting physics is expected
to emerge from case $\#3$, where both resonances become crucial
in determination of the neutrino behavior.
We have obtained some preliminary results for various values of the parameters,
using the data for the ${^{37}} \! Cl$ detector. We find that upon inclusion
of the third flavor, certain
parameters may be chosen such that the ``large mixing region" can be
widened.  To see the magnitude of the widening, we compare the width of
the region for 2 flavors to the width using 3 flavors (with two choices
of $\tetau$).  For the three-flavor model ({\it i.e.} $\tetau \neq 0$),
we choose $\df_{31} = 10^{-15}$.

\begin{table}[h]
\caption{{\it Width of Allowed Parameter Space for Large Mixing Region
as a function of $\sin^2 \tetau$}}
\vspace{7mm}
\begin{tabular}{|c|c|} \hline
$\sin^2 \tetau$ & Width ($\sin^2 2 \temu$) \\ \hline\hline
0    & 0.60  -  0.89 \\ \hline
0.01 & 0.65  -  0.94 \\ \hline
0.05 & 0.64  -  0.93 \\ \hline
0.10 & 0.63  -  0.93 \\ \hline
0.40 & 0.57  -  0.91 \\ \hline\hline
\end{tabular}
\end{table}
\vspace{10mm}

{}From table 1 we see that the known parameter space overlap for the
$\nu_e \ra \nu_\mu$ transitions is shifted and expanded somewhat with the
addition of  the $\nu_{\tau}$.
Further study of the adiabatic behavior
of the neutrinos with respect to the SNP may yield a nice set of parameters
which maximize the allowed region.  Also, we may narrow down our possible
choices of violation parameters and mixing angles through
a comparison of detected to theoretically calculated
${^8}B$ neutrino spectra.  {\it Fig. 1} shows the calculated flux for two
of the parameter sets listed in Table 1, for a fixed SNU rate of
${^8}B$ neutrinos (i.e. $C = 1.38$ SNU).  Note the subtle shifts in shape
of each reduced spectrum for the lower half of the energy range, {\it i.e.}
below approximately $6$ MeV ({\it Fig.1}), as well as the flattening of
the flux curve for increasing $\tetau$ in {\it Fig.2}.  These differences could
possibly be differentiated by present-day or near-future detectors (see
{\it e.g.} some discussions in \cite{bald,msw2,min}).

	A VEP analysis incorporating all three generations
of neutrinos seems to yield a greater mobility in parameter space.
Such an analysis is quite timely, due to the advent of such new ventures
as the Sudbury Neutrino Observatory, for more precise measurements of
neutrino fluxes can help determine the right set of parameters.  As
we can see from {\it Figs. 1}\, and {\it 2}, there is greater range of
parameters from which to choose such that we can exactly match a flux
curve.  Since SNO will be able to detect the full solar neutrino flux
(flavor-independent via neutral current interactions),  studies of such
spectra could also help determine the reduction mechanism at work, and
hence possibly distinguish between a VEP- or MSW-based phenomenon.
Also, an analysis of the non-adiabatic transitions is
currently underway, to gauge what effect this will have on the
full parameter space range.  Since the small mixing region is
dominated by  highly non-adiabatic transitions, it is believed
that this could represent a larger portion of the expanded
three-flavor space.

\bigskip\noindent
{\bf Acknowledgements}

J.M. would like to thank Rob Malaney for various insightful
discussions. This work was supported by the Natural Sciences and Engineering
Research Council of Canada.

\pagebreak

{\large\bf Figure Captions} \\
\\

{\bf Fig. 1}: Plot of reduced $^8B$ neutrino fluxes ($cm^{-2}s^{-1}MeV^{-1}$)
v.s.\
neutrino energy ($MeV$) for fixed counting rate $C_{^8B} = 1.38$ SNU
showing variation of spectrum shape with choice of parameters.  Upper (large)
flux is unreduced SSM $^8B$ flux from \cite{bah3}, while reduced flux curves
for 3-flavor model are calculated with $\df_{21} = 10^{-16}$, $\df_{31} =
10^{-15}$, and \\
\begin{itemize}
\item $sin^2 2 \temu = 0.685$ ; $sin^2 \tetau = 0.40$  (dotted)
\item $sin^2 2 \temu = 0.70$ ; $sin^2 \tetau = 0.05$  (solid)
\end{itemize}
These are compared with predicted 2-flavor flux ($\df_{31} \ge 10^{-12}$), with
\begin{itemize}
\item $sin^2 2 \temu = 0.697$ ; $sin^2 \tetau = 0$  (dashed)
\end{itemize}
\vspace{6mm}
\begin{center}
\rule{70mm}{.5mm}
\end{center}
\vspace{6mm}
{\bf Fig. 2}: Plot of reduced $^8B$ neutrino fluxes v.s.\ neutrino energy for
fixed counting rate as would be detected by $^{37}Cl$ experiments
({\it i.e.} ${^8}B+{^7}Be$), $C_{Cl} = 2.00$ SNU.
Reduced flux curves for 3-flavor model are calculated with $\df_{21} =
10^{-16}$,
$\df_{31} = 10^{-15}$, and \\
\begin{itemize}
\item $sin^2 2 \temu = 0.63$ ; $sin^2 \tetau = 0.40$  (dotted)
\item $sin^2 2 \temu = 0.693$ ; $sin^2 \tetau = 0.05$  (solid)
\end{itemize}
These are compared with predicted 2-flavor flux ($\df_{31} \ge 10^{-12})$, with
\begin{itemize}
\item $sin^2 2 \temu = 0.701$ ; $sin^2 \tetau = 0$  (dashed)
\end{itemize}
\vspace{6mm}
\begin{center}
\rule{70mm}{.5mm}
\end{center}
\vspace{6mm}

\end{document}